\documentclass[12pt]{iopart}
\pdfoutput=1
\usepackage{amsfonts}
\usepackage{graphicx}
\usepackage{ulem}
\usepackage{xcolor}
\usepackage{subfigure}
\usepackage{textcomp}
\usepackage{cite}
\usepackage{float}
\usepackage{soul}
\usepackage{stackrel}
\usepackage{hyperref}

\definecolor{dgreen}{rgb}{0,0.7,0}

\expandafter\let\csname equation*\endcsname\relax
\expandafter\let\csname endequation*\endcsname\relax
\usepackage{amsmath,amssymb}

\pdfoutput=1
\usepackage{esint}

\usepackage{color}
\definecolor{dgreen}{rgb}{0,0.7,0}

\def\bea{\begin{eqnarray}}
\def\eea{\end{eqnarray}}

\def\nn{\nonumber}

\newcommand{\aref}[1]{\ref{#1}}%

\renewcommand{\eref}[1]{Eq.~(\ref{#1})}%
\renewcommand{\fref}[1]{Fig.~\ref{#1}} %
\renewcommand{\sref}[1]{Sec.~\ref{#1}}%

\begin{document}

\title{Queues with resetting: a perspective}

 \author{Reshmi Roy$^{1,2}$, Arup Biswas$^{1,2}$, Arnab Pal$^{1}$}

 \address{$^{1}$The Institute of Mathematical Sciences, CIT Campus, Taramani, Chennai 600113, India \& Homi Bhabha National Institute, Training School Complex, Anushakti Nagar, Mumbai 400094, India}
 \address{$^{2}$These authors contributed equally to this work}

\ead{reshmiroy@imsc.res.in,arupb@imsc.res.in,arnabpal@imsc.res.in}
\vspace{10pt}

\begin{abstract}
Performance modeling is a key issue in queuing theory and operation research. It is well-known that the length of a queue that awaits service or the time spent by a job in a queue depends not only on the  service rate, but also crucially on the fluctuations in service time. The larger the fluctuations, the longer the delay becomes and hence, this is a major hindrance for the queue to operate efficiently. Various strategies have been adapted to prevent this drawback. In this perspective, we investigate the effects of one such novel strategy namely resetting or restart, an emerging concept in statistical physics and stochastic complex process, that was recently introduced to mitigate fluctuations-induced delays in queues. In particular, we show that a service resetting mechanism accompanied with an overhead time can remarkably shorten the average queue lengths and waiting times. We examine various resetting strategies and further shed light on the intricate role of the overhead times to the queuing performance. Our analysis opens up future avenues in operation research where resetting-based strategies can be universally promising.

\end{abstract}

\section{Introduction} \label{introduction}
Queuing theory is usually considered to be a branch of operations research that
mathematically studies the formation, function and other aspects of waiting lines that stretch in front of a service station \cite{Adan-book,Cohen-book,Haviv-book,Newell-book}. 
Queues are ubiquitous in nature and they appear in a wide range of applications 
ranging from supermarkets, banks, call-centers \cite{Call1,Call2}, telecommunications \cite{Tele1,Tele2}, airplane boarding \cite{Plane1,Plane2,Plane3}, computer systems \cite{CSqueue1,Mor-book}, emergency services, transport phenomena\cite{Transport2,Transport4,Transport7} to gene expression \cite{Gene1,Gene2,Gene4,Gene5,Gene6} enzymatic and metabolic pathways \cite{enzymatic3,enzymatic4,enzymatic5,enzymatic6,enzymatic8}. Each set-up of a queue has its unique working principle. For example, a teller in the bank or a supermarket may work more or less at a constant rate but this can not be said for the computer servers or living entities like genes or enzymes which may often display more fluctuations in service time \cite{Large_fluc2,Large_fluc3,Large_fluc4,Large_fluc6}. In fact, it is now a well-established fact that the efficiency of a queue depends not only on the rate of the server but it is also extremely sensitive to the stochastic fluctuations in service times. As a result, these fluctuations have profound consequences and quite often they render acute backlogs and delays in queues stalling the work-conditions \cite{Mor-book}.

There has been a persistent strive to tailor generic strategies that can control and mitigate the adverse effect caused by stochastic service time fluctuations in queues especially the ones that encounter heavy tailed workloads \cite{Whitt-2000}. Various scheduling policies have been developed eg, small jobs are being served first by the server instead of first-come-first-serve. Although this policy can be proven optimal under certain conditions, it is also criticized due to its lack of fairness \cite{Mor-book}. Notably, these policies are applicable to queues where the source of service time fluctuations is rather extrinsic i.e, it depends on the variability in the job sizes or the numbers of items as in the supermarkets. However, these policies are not well-equipped to deal with situations where fluctuations in service times are intrinsic to the server itself. This is indeed the case for stochastic optimization algorithms, genes or the enzymes where stochastic fluctuations are intrinsic to the server \cite{SO,Large_fluc1,Large_fluc5}. Naturally, these policies turn out to be inadequate to be implemented for such scenarios and thus, novel approaches are very much in need.

In a recent work \cite{bonomo_pnas_2022}, we proposed a novel approach, based on resetting or restart, to mitigate the problems caused by service time fluctuations in queuing systems. It was shown that the length of a queue can be significantly shortened by a simple service resetting policy. To understand the resetting mechanism, consider an arbitrary stochastic dynamical process which completes a task in some random time. However, the process can be restarted (i.e, started from scratch) intermittently before the completion of the task and thus it has to begin completely anew and repeat the same task \cite{Restart1,Restart2,Review,IP,PalReuveniPRL17,branching,Restart-Search3,recordage,CS3,Pal-potential,expt,expt-2,expt-3}. This procedure repeats itself until the process reaches completion. The completion time will thus depend on the details of the underlying process and the resetting protocol. In a similar vein, consider a single server queue where the server has a task of completing one job at a time. This completion takes a random time -- moreover, the source of stochastic fluctuations is considered to be intrinsic. To implement resetting, imagine this server being stopped intermittently and  then restarted -- thus, jobs whose
service has been reset are now assigned fresh service times.

 At a first glance one might wonder why restarting from scratch can expedite the completion of a complex random process. Indeed, this has been a quest in statistical physics and stochastic process for the last decade where resetting has been shown to systematically eliminate errand trajectories and find alternative pathways that can be avoid potential obstacles \cite{Restart1,Restart2,Review,IP,PalReuveniPRL17,branching,Pal-book,Satya-refractory,obstacle,nld,channel,Sandev}. Unveiling new such trajectories can often fasten the completion in particular when the resetting-free underlying processes possess search time with large fluctuations. This particular observation is quite intriguing as it states that resetting can overturn the high uncertainty due to large fluctuations in the underlying search time, thus converting a drawback into a favorable advantage. No wonder, why such resetting based strategies have been proven to be extremely useful in a wide range of search processes spanning from stochastic optimization \cite{algorithm-1,algorithm-2,algorithm-3}, first passage processes\cite{Restart-Search1,interval,Das1,Campos,Yin}, home-based foraging \cite{Palprr20}, chemical reactions \cite{ReuveniEnzyme1,Gating}, income dynamics \cite{Stojkoski-income-1,Income-2} to transport over the last few years. 

The goal of this perspective is to first introduce the topics ``queues'' and ``resetting'', review the formulation of a single-server queue with instantaneous service resetting and to bring new insights when the service resetting is not instantaneous but involves an overhead or buffer time. We will discuss the ramifications of this overhead time and its interplay with the service time on the length of the queue or the waiting time. A pedagogic approach has been taken keeping the non-expert readers in mind.

The rest of the paper is structured in the following way. In \sref{model} ``Preliminaries", we provide a brief overview of the M/G/1 queuing system with general service and Markovian job arrival -- this will serve as the prototype model of a queue in this paper. We discuss the Pollaczek-Khinchin formula which gives an estimation of the mean number of jobs in the steady state of the M/G/1 queue, emphasizing the dependency of the latter to the service time variability. In \sref{Sec 3}  ``Service with resetting'', we formulate the M/G/1 queuing model with service resetting where we associate a refractory time to the overall process. In effect, this `modified' queue is similar to the standard queue with a compounded service time that depends on the resetting and overhead time. With this mapping, we can immediately use the Pollaczek-Khinchin formula in the M/G/1 queue with a modified service time to compute the mean number of jobs. In \sref{poisson resetting sec} ``M/G/1 Queues with Poissonian service resetting", we discuss the effects of Poissonian service resetting (i.e., resetting with a constant rate) on the mean number of jobs in the queue where the calculations simplify in multi-fold. In \sref{Service at ORR} ``Service at an optimal resetting rate'', we study the mean number of jobs when the service is restarted at an optimal resetting rate under Poissonian resetting. We move on to demonstrate how resetting can reduce the number of jobs depending on the variability of the overhead time. In \sref{Application} ``Application", we illustrate how the general formalism developed so far can be used for the log-normal service time distribution and overhead times with different variability. We conclude in \sref{Discussion} ``Discussion and Summary" with a summary and future perspective. Appendices provide detailed derivations and other technical results to keep the paper self-contained.

In what follows, we use the notations $f_X(t)$, $\langle X \rangle$, $\sigma^2(X)$, $\widetilde{X}(s)~\equiv~\langle e^{-sX}\rangle$ and $CV_X$ to denote, respectively, the probability density
function, expectation, variance, Laplace
transform and coefficient of variation of a non-negative random variable $X$.

\section{Preliminaries}\label{model}
Let us consider a single line queuing system where a server serves one job at a time and jobs await to be served in a first come first serve basis. Such a queue is often represented as M/G/1 queue in Kendall's notation where the notation M stands for the Markovian or memory-less arrival of jobs. Here, we assume that the jobs arrive according to a Poisson process with rate $\lambda$ and G stands for the service time of jobs, which can be drawn from a general distribution. We indicate this service time random variable as $S$ which is distributed according to $f_S(t)$. The last notation 1 simply indicates that one job should be served at a time. Notably, we assume that the server needs to wait some overhead time following a reset or a service. This is quite common in a computer software or algorithm where a buffer time is required to initialize and reload the program. Similar situations also appear in natural systems such as the chemical reaction or facilitated diffusion. We denote the overhead time by the random variable $S_{on}$. Therefore, the total service time  of the underlying process ($S_u$) is defined by the sum of $S$ and $S_{on}$ with the corresponding density $f_{S_u}(t)$. The service rate
$\eta$ of this process can be defined as $\eta=1/\langle S_u \rangle$. Furthermore, one can define the utilization parameter $\rho=\lambda/\eta$ that signifies the
fraction of time the server works in steady state. To attain the steady state of the system the arrival rate of jobs $\lambda$ must be less than the effective service rate $\langle S_u \rangle^{-1}$ so that $\rho<1$. Otherwise, for $\rho>1$, the number of jobs in the queue will blow up and the length of the queue will increase indefinitely. 

The state space of the M/G/1 queue is denoted
by the set $N = {0, 1, 2, 3,...}$, where the value of $N$ corresponds to the number of jobs in the queue, including the one being served. This number fluctuates in time as the arrival and service are random processes. One key observable in the queuing theory is the mean number of jobs $\langle N \rangle$ 
in the system (queue+server) which is given by the famous Pollaczek-Khinchin formula in the steady state \cite{Mor-book}. 
\begin{align}
    \langle N \rangle
=\frac{\rho}{1-\rho}+\frac{\rho^2}{2(1-\rho)}\left( CV_u^2-1  \right)~,
\label{PK-1}
\end{align}
where $\rho$ is the utilization and 
\begin{align}
 CV_u^2=\frac{\sigma^2 (S_u)}{\langle S_u \rangle^2},
 \end{align}
is the squared coefficient of variation or the variability in total service time of the underlying process. Several comments can be made here. First, note that the mean number of jobs increases monotonically as a function of the utilization $\rho$ and it diverges as $\rho \to 1$ leading to ``piling up'' of jobs. Secondly, $\langle N \rangle$ is found to be highly sensitive to the variability in the service time. Namely, if $CV_u<1$, the second term on the RHS in \eref{PK-1} becomes negative leading to shorter queues. On the other hand, if the service time has large fluctuations namely $CV_u>1$, the second term adds positive contribution to $\langle N \rangle$ leading to longer queues. It is thus evident that service time fluctuations are central to
the behavior of the M/G/1 queue, and their effect in other queuing systems can also be anticipated in a similar way.

The mean waiting time $\langle T \rangle$ of a job in the queue, i.e., the time elapsed from arrival to the end of service, is proportional to $\langle N \rangle$ via Little's law  $\langle T \rangle=\lambda^{-1}~\langle N \rangle$ \cite{Mor-book}. This results in
\bea
\langle T \rangle=
\frac{1/\eta}{1-\rho}+\frac{\rho/\eta}{2(1-\rho)}\left( CV_u^2-1  \right)~,
\label{PK-2}
\eea
which again crucially depends on fluctuations in service time similar to the mean number of jobs. The fluctuations in service time can both be intrinsic and extrinsic to the server. If a server serves jobs of different
sizes with constant rate, then the service time is extrinsic to the server and it will be dictated by the job size. As in supermarkets, the service time for the customers at the billing counter is determined by the number of items each customer buys. On the other hand, the fluctuation in the service time can also be intrinsic to the server itself.
The catalytic reaction can be an example of this where to catalyze chemical reaction an enzyme takes different time between turnover cycle though the substrate and product molecule are chemically identical. We refer to \cite{bonomo_pnas_2022} for a detailed discussion on the origin of service time fluctuations in queues.

\begin{figure}
    \centering
    \includegraphics[scale=1.4]{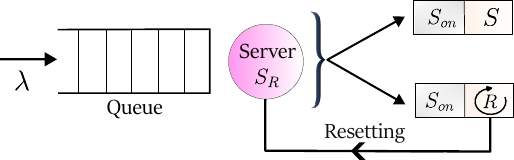}
    \caption{Schematic of a queuing system under resetting. Jobs arrive at the queue with a rate $\lambda$ and they are being served at the service station according to a first-come, first-served policy. The server has two components: a service time $S$ followed by an overhead time $S_{on}$. It can complete a task in time $S_u=S+S_{on}$ prior to resetting (upper branch). Otherwise, resetting can occur in time $R$ (lower branch), following which the service is renewed. The process repeats by itself until the task is completed which is possible only via the upper branch.}
    \label{schematic}
\end{figure}

\section{Service with resetting}
\label{Sec 3}
When service time fluctuations are intrinsic to the server in a queue, a service resetting can be implemented in the following way. Again we recall the M/G/1 queue where the jobs are being served one at a time. Consider the server that starts at time zero and, if allowed to take place
without interruptions, completes after a random time $S_{on}+S$. The
service, however, is restarted at some random time $R$ following which service renews. Denoting the 
random service time of the compound process by $S_R$
it can be seen that
\begin{equation}
\begin{array}{l}
S_{R}=S_{on}+\left\{ \begin{array}{lll}
S &  & \text{if ~~}S<R\text{ ,}\\
 & \text{ \ \ }\\
R+S_R'&  & \text{if~~ }R\leq S\text{ ,}
\end{array}\right.\text{ }\end{array}
\label{renewal-1-main}
\end{equation} 
where $S_{on}$ is a random time drawn from a general distribution which accounts for the time delay that may occur prior to either service completion or resetting and $S_{R}'$ is an independent and identically distributed copy of $S_{R}$. To understand this equation, observe that when service occurs before restart, $S_R = S_{on}+S=S_u$. However, if service is restarted at a time $R \leq S$, then a new service time $S$ is drawn, and service restarts following the overhead time $S_{on}$. In that case, we simply have $S_R = S_{on}+R + S_R'$. Thus, it can be seen that \eref{renewal-1-main} forms a renewal structure. As shown in \cite{bonomo_pnas_2022}, here the service mechanism can be understood as a first passage process which is intermittently subjected to the resetting strategy. A comprehensive framework for first passage under resetting was developed in \cite{ReuveniPRL16,PalReuveniPRL17}. We review this method partially in the (\aref{appa})  and only present the key results here. To this end, note that \eref{renewal-1-main} can be recast in the following way
\begin{align}
    S_R=S_{on}+\text{min}(S,R)+I(R\leq S) S_R',
    \label{renewal-re}
\end{align}
where $\text{min}(S,R)$ is the
minimum of $S$ and $R$ and $I({R \leq S })$ is an indicator random variable which takes the value one when $R \leq S$ and
zero otherwise. \eref{renewal-re} can be used to compute the moment generating function of $S_R$ which we keep to the Appendix. The first and second moments however can be computed directly by taking expectations on both sides of \eref{renewal-re}. Performing the averages, one finds
\begin{align}
\langle S_R  \rangle &= \frac{\langle \text{min}(S,R) \rangle +\langle S_{on} \rangle}{Pr(S<R)} ,
\label{first-mom} \\
\langle S_R^2  \rangle & = \frac{\langle (\text{min}(S,R)+S_{on})^2\rangle }{Pr(S<R)}+\frac{2Pr(R\le S)(\langle R_{min}\rangle +\langle S_{on} \rangle)\left(\langle \text{min}(S,R) \rangle + \langle S_{on}\rangle\right)}{Pr(S<R)^2}   \label{sec-mom} 
\end{align}
where $Pr(S<R)$ is the probability of service being completed prior to restart, and $R_{\text{min}}=\{R|R<S\}$
stands for the conditional restart time given that restart occurred prior to service. Finally, recall that the variance in the service time is given by $\sigma^2(S_R)=\langle S_R^2 \rangle -\langle S_R \rangle ^2$. Given the distribution of the resetting times and service times, it is straightforward task to compute both the moments in \eref{first-mom} and \eref{sec-mom}, as we will show explicitly in the next section. 


\section{M/G/1 Queues with Poissonian service resetting} \label{poisson resetting sec}
There are numerous possible ways in which service and resetting mechanisms can mix and match. One such resetting mechanism namely the Poissonian resetting has been extensively studied in the recent past  \cite{Restart1,Restart2,Restart-Search1,Restart-Search2,ReuveniPRL16,PalReuveniPRL17, interval,Das1,expt, Restart-Search3,Pal-potential}. As the name suggests, here the number of resetting events in a given time interval is distributed according to the Poisson distribution. More on the technical ground, if the resetting occurs at a rate $r$, the mean number of resetting events in time $t$ is given by $rt$ and the resetting time $R$ is drawn from an exponential distribution i.e., $f_R(t)=re^{-rt}$. The mean and second moment of the service time can then be derived using Eq.s (\ref{first-mom}) and (\ref{sec-mom}) (see (\aref{appb}) for the derivation)
\begin{align}
    \langle S_r \rangle &= \frac{1-\widetilde{S}(r)+r \langle S_{on} \rangle}{r \widetilde{S}(r)}, \label{mean under restart} \\
\langle S_r^2 \rangle &= \frac{2 r \frac{d \widetilde{S}(r)}{dr}(1+r \langle S_{on}\rangle)+2(1-\widetilde{S}(r))(1+r \langle S_{on} \rangle)^2+
r^2 \widetilde{S}(r)\langle {S_{on}^2} \rangle}{r^2 \widetilde{S}(r)^2},
\label{secmom under restart}
\end{align}
where $\widetilde{S}(r)=\int_0^\infty~dt~ e^{-rt}~f_S(t)$ is the Laplace transform of the service time $S$, evaluated at the restart rate $r$. The utilization of this queue is then given by $\rho_r=\lambda\langle S_r \rangle$, and the squared coefficient of variation of the service time is  $CV_r^2=\sigma^2(S_r)/\langle S_r \rangle^2$. 

Under resetting mechanism, one notices that the queue service time is now modified to $S_r$. Henceforth, one can replace $\rho$ with $\rho_r$, and $CV_u^2$ with $CV_r^2$, in \eref{PK-1} to compute the mean queue length under resetting. This yields  
\bea
\langle N_r \rangle=\frac{\rho_r}{1-\rho_r}+\frac{\rho_r^2}{2(1-\rho_r)}\left( CV_r^2-1  \right)~.
\label{PK-1-r}
\eea
Similarly, the mean waiting time $\langle T_r \rangle$ in the system can be derived from Little's law \cite{Nelson-book,Mor-book}, yielding an analogous result to \eref{PK-2}. 
To demonstrate the effect of resetting in a M/G/1 queue in the presence of a overhead time, we consider Fig. \ref{fig1}(a) as an illustrative example (details will be discussed in Sec. \ref{Application}). Commencing from a service time distribution, service resetting has been introduced and
the mean service time $\langle S_r \rangle$ is plotted against the resetting rate $r$ for various overhead time distribution. We note that $\langle S_r \rangle$ initially decreases, obtaining a minima at an optimal resetting rate $r^*$ before increasing further, as evident from the figure. The emergence of the optimal resetting rate $r^*$ is quite noteworthy as resetting does not only lower the mean service time but also renders a global minimum time. The next section is dedicated to a detailed analysis of $r^*$ and how it is connected to the generic service and overhead time.

\section{Service at an optimal resetting rate} \label{Service at ORR}
The optimal resetting rate that minimizes the mean service time can be obtained by setting
\begin{align}
 \frac{d\langle S_r \rangle}{dr}\bigg|_{r=r^*}=0.   
 \label{ORR-condition}
\end{align}
Substituting the expression for 
$\langle S_r \rangle$ from equation (\ref{mean under restart}) into \eref{ORR-condition}, we find
\bea
\widetilde S(r^*) (\widetilde S(r^*)-1)-{(r^*)\widetilde S^\prime(r^*)}(1+r^*\langle S_{on}\rangle) =0,
\eea
where $\widetilde S^\prime(r)$ refers to the derivative of $\widetilde S(r)$ as a function of $r$. Using $\widetilde S^\prime(r)$ from the above expression in \eref{secmom under restart} for the second moment $\langle S_{r^*}^2 \rangle$, and simplifying further one arrives at the following  relation for the variability of an optimally restarted process
\cite{ReuveniPRL16}
\begin{align}
    CV_{r^*}=\frac{\sigma(S_{r^*})}{\langle S_{r^*}\rangle}= \sqrt{1+\frac{\langle S_{on}\rangle^2}{\widetilde S(r^*)\langle S_r^*\rangle^2}\left(CV^2_{on}-1 \right)},
\label{CV*}
\end{align}
where $CV_{on}=\frac{\sigma(S_{on})}{\langle S_{on}\rangle}$ is the variability of the overhead time and furthermore we have rewritten $CV_u$ in terms of the $S_{on}$-metrics in the following way
\begin{align}
    CV_u=\sqrt{\frac{\sigma^2(S)+\sigma^2(S_{on})}{\langle S+S_{on} \rangle^2}}.
\label{cvu}
\end{align}
The relation in \eref{CV*} is completely universal since it does not depend on the specific choice of the underlying service time and the overhead time. Our next goal is to understand how the mean queue length in Pollaczek-Khinchin formula would change with optimal resetting.

The mean length of the queue for the optimally restarted process is then given by the Pollaczek-Khinchin formula in \eref{PK-1-r} with the substitution $r \to r^*$. This results in
\begin{align}
    \langle N_{r^*} \rangle=\frac{\rho_{r^*}}{1-\rho_{r^*}}+\frac{\rho_{r^*}^2}{2(1-\rho_{r^*})}\left( CV_{r^*}^2-1  \right).
\end{align}
Now we can use the universal relation (\ref{CV*}) for the optimally restarted process in above to find
\begin{align}
       \langle N_{r^*} \rangle&=\frac{\rho_{r^*}}{1-\rho_{r^*}}+\frac{\lambda^2}{2(1-\rho_{r^*})} \frac{\langle S_{on}\rangle^2}{\widetilde S(r^*)}\left( CV_{on}^2-1  \right), \label{NrCVon}
\end{align}
where we have also used the fact that $\rho_{r^*}=\lambda\langle S_{r^*} \rangle$. \eref{NrCVon} suggests that the mean queue length depends crucially on the variability of the overhead time. In the following section, we identify a few such different cases based on $CV_{on}$. 

 So far in the analysis, we have assumed that there exists a finite optimal resetting rate. But it is worthwhile to ask under what conditions this is guaranteed. Skipping details of the proof from \aref{appc}, we note that the  general criteria that ensures the existence of an optimal $r^*$ reads
\begin{align}
    CV^2_u>1+ \frac{\langle S_{on} \rangle^2}{\langle S_u \rangle^2}(CV_{on}^2-1),
    \label{cvu_cond}
    \end{align}
 which again crucially depends on $CV_{on}$.

\subsection{Resetting with no overhead}
We first recap the scenario when there is no overhead time in the system. Thus, the service restarts immediately. This was well studied in \cite{bonomo_pnas_2022}. 
In this case, $S_u=S$ and thus \eref{mean under restart} and \ref{secmom under restart} simply reduce to \cite{bonomo_pnas_2022}
\begin{align}
    \langle S_r \rangle &= \frac{1-\widetilde{S}(r)}{r \widetilde{S}(r)},  \\
\langle S_r^2 \rangle &= \frac{2 r \frac{d \widetilde{S}(r)}{dr}-\widetilde{S}(r)+1} {r^2 \widetilde{S}(r)^2},
\end{align}
Moreover, the universal relation (\ref{CV*}) simply becomes $CV_r^*=1$. As a result, the mean queue length turns out to be
\begin{align}
\langle N_{r^*}\rangle=\frac{\rho_{r^*}}{1-\rho_{r^*}},
 \end{align}
which satisfies the following inequality $\langle N_{r^*}\rangle < \langle N \rangle$. Thus, the mean number of jobs in the queue can be reduced by resetting service at an optimal rate. Moreover, the criterion for a finite $r^*$ reduces to $CV_u>1$. We refer to \cite{bonomo_pnas_2022} for more details.

\subsection{Overhead time with $CV_{on}<1$}
\label{cvl1}
We now consider a scenario when the overhead times are sampled from a distribution which is narrowly dispersed i.e., $CV_{on}<1$. Denoting the 
queue length by $ \langle N_{r^*}^I \rangle$, one finds
    \begin{align}
       \langle N_{r^*}^I \rangle&=\frac{\rho_{r^*}}{1-\rho_{r^*}}+\frac{\lambda^2}{2(1-\rho_{r^*})} \frac{\langle S_{on}\rangle^2}{\widetilde S(r^*)}\left( CV_{on}^2-1  \right)
       \label{nr1}
\end{align}
where the second term in the above equation gives a 
negative contribution so that 
$\langle N_{r^*}^I \rangle<\frac{\rho_{r^*}}{1-\rho_{r^*}}$. The general condition in Eq. (\ref{cvu_cond}) suggests that for the case $CV_{on}<1$, one should have $CV_u>1$. This is a sufficient (but not necessary) condition that guarantees that a finite $r^*$ should exist.

Now, if a non-zero finite $r^*$ exists,
we should have $\langle S_{r^*} \rangle<\langle S_u \rangle$ which essentially implies $\rho_{r^*}<\rho$ where recall $\rho =\lambda \langle S_u \rangle$ and $\rho_r=\lambda \langle S_r \rangle$.
 Finally, noting that $\frac{\rho}{1-\rho}$ is a monotonically increasing function of $\rho$, it becomes explicit that $\frac{\rho_{r^*}}{1-\rho_{r^*}}< \frac{\rho}{1-\rho}$. Collecting all the pieces together we arrive at the following hierarchical inequality
\begin{align}
    &\langle N_{r^*}^I \rangle<\frac{\rho_{r^*}}{1-\rho_{r^*}}<\frac{\rho}{1-\rho}\le \frac{\rho}{1-\rho}+\frac{\rho^2}{2(1-\rho)}\left( CV_u^2-1  \right)=\langle N \rangle,
    \label{con1}
\end{align}
which will always holds as long as $CV_u>1$ and thus service resetting will certainly help to alleviate the queue.  
Since the criterion $CV_u>1$ is not a necessary one, a finite optimal resetting rate $r^*$ may exist (resulting in a reduction in the queue length) even before $CV_u=1$ (see \aref{cvl1} for additional discussion). This analysis effectively shows that service resetting can reduce the mean queue length even in the presence of finite overhead times.

\subsection{Overhead time with $CV_{on}=1$}
Next, we turn our attention to the marginal case when $CV_{on}=1$. In this case, the second term in \eref{NrCVon} vanishes and the mean queue length, denoted by $\langle N_{r^*}^{II} \rangle$, becomes
\begin{align}
    \langle N_{r^*}^{II} \rangle=\frac{\rho_{r^*}}{1-\rho_{r^*}}. \label{nr2}
\end{align}
A simple manipulation then shows
\begin{align}
      &\langle N_{r^*}^{II} \rangle=\frac{\rho_{r^*}}{1-\rho_{r^*}}\le \frac{\rho}{1-\rho}\le \frac{\rho}{1-\rho}+\frac{\rho^2}{2(1-\rho)}\left( CV_u^2-1  \right)=\langle N \rangle,
      \label{con2}
\end{align}
where we have used the similar line of rationale as in the previous case and further argued that $CV_u>1$ for the existence of a finite $r^*$ (see \aref{cv1} for additional discussion). 

\subsection{Overhead time with $CV_{on}>1$}
Finally, we consider a scenario where the overhead times are drawn from a distribution which is broadly dispersed so that $CV_{on}>1$. In this case, we have
\begin{align}
      \langle N_{r^*}^{III} \rangle&= \frac{\rho_{r^*}}{1-\rho_{r^*}}+\frac{\lambda^2}{2(1-\rho_{r^*})} \frac{\langle S_{on}\rangle^2}{\tilde S(r^*)}\left( CV_{on}^2-1  \right), \label{nr3}
\end{align}
where the second term is strictly positive. However, in this case, $CV_u>1$ condition does not guarantee $ \langle N_{r^*}^{III} \rangle < \langle N \rangle$ as was done in the previous subsections. However, one can do a more careful analysis to show that there exists a sufficient (not necessary) condition that can reassure the inequality $ \langle N_{r^*}^{III} \rangle < \langle N \rangle$. This modified criterion reads (see \aref{cvg1} for details) 
\begin{align}
    CV_u^2>1+\left(\frac{\langle S_{on}\rangle}{\langle S_u \rangle}\right)^2\frac{1}{\Tilde{S}(r^*)}(CV_{on}^2-1). \label{cvon_cr}   \end{align}
Since $CV_u$ is a control parameter in this problem, we can impose the above condition to see a reduction in the mean queue length under service resetting.

\section{Application} \label{Application}
To demonstrate the power of our approach, we consider a M/G/1 queue whose service times are distributed according to a log-normal distribution -- a well-known service time distribution in the
queuing literature \cite{log-normal-callcenter1,log-normal-callcenter2}. We will study the effect of resetting on this server for different overhead time distributions. We start with the following form of the log-normal distribution
\bea
f_S(t)=\frac{1}{\sqrt{2\pi} \alpha t}~e^{-\frac{(\ln t-\mu)^2}{2\alpha^2}}, \label{log normal dist}
\eea
for $t>0$, where $\mu \in (-\infty, \infty)$ and $\alpha>0$. The mean and variance of the service time $\langle S \rangle$ in this case are given by
\bea
&\langle S \rangle&~=~e^{\mu + \frac{\alpha^2}{2}}, \label{mean log normal} \\ 
&\sigma(S)&~=~\left(e^{\alpha^2} -1\right)e^{2\mu+\alpha^2}, \label{var log normal}
\eea
such that 
\bea
CV^2=e^{\alpha^2} - 1~, \label{CV log normal}
\eea
which is independent of $\mu$. In the previous sections, we have discussed the general results when the overhead time is drawn from some arbitrary distributions and furthermore illustrated the role of its variability $CV_{on}$. In what follows, we will focus on three representative cases with different $CV_{on}$ but keeping $\langle S_{on} \rangle$ fixed. Finally, we discuss the scenario when the latter condition is also relaxed.

\begin{figure}[t]
\includegraphics[scale=1.23]{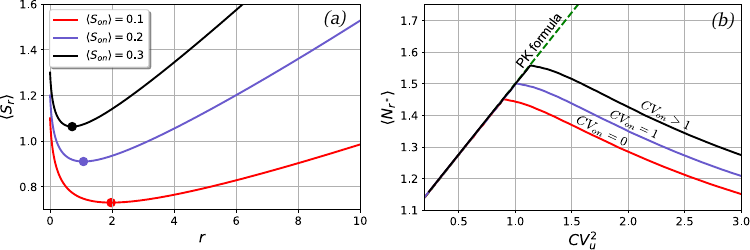}
\caption{Panel (a) The mean service time $\langle S_r \rangle$ from Eq. (\ref{mean under restart}) as a function of the resetting rate for different $\langle S_{on} \rangle$. The service time $S$ is sampled from the log-normal distribution whose density is given by \eref{log normal dist}. Here, we have set $\langle S \rangle = 1$, and $\alpha = 1.5$. The circles show the optimal resetting rate $r^*$ for different $ \langle S_{on} \rangle$, where $\langle S_{r} \rangle$ attains consecutive minima. Panel (b) shows the mean queue length at optimal resetting i.e., $\langle N_{r^*}  \rangle$ as a function of the squared $CV_u$. Note that $\langle S \rangle$ is fixed at unity and $\alpha$ is varied to control the stochastic fluctuation in $CV_u$ via \eref{cv_deterministic_equation}, \eref{cv_exponential_equation}  and \eref{cv_weibull_equation} respectively. The overhead time is taken from three different distributions: deterministic ($CV_{on}=0$), exponential ($CV_{on}=1$) and Weibull ($CV_{on}>1$) while we have 
set $\langle S_{on} \rangle=0.5$ fixed for all the plots. The Pollaczek–Khinchin formula (shown by the dashed line) gives the familiar linear dependence of \eref{PK-1} when $r^*=0$. However, as we vary $CV_u$ further, the optimal resetting rate $r^*$ becomes finite and thus the plots for $\langle N_{r^*}  \rangle$ deviate from the same with $r^*=0$. The transition points (where the deviation occurs) can be corroborated with the theory as explained in the main text. Indeed, resetting at an optimal rate can significantly shorten the mean queue length for any $S_{on}$ with $CV_{on} \{<1,=1,>1 \}$. In all the plots, we have set $\lambda=0.4$.}
\label{fig1}
\end{figure}

\subsection{Case I: $CV_{on}<1$}
 We start with the case when $CV_{on}=0<1$ i.e., the overhead time distribution is sharply peaked around its mean $\langle S_{on} \rangle$ so that $f_{S_{on}}(t)=\delta(t-\langle S_{on} \rangle)$. In this case, variability of the 
underlying process can be expressed as
\begin{align}
    CV^2_u=\frac{e^{\alpha^2}-1}{(1+ \langle S_{on}\rangle)^2}.
    \label{cv_deterministic_equation}
\end{align}
In Fig. \ref{fig1}(a), we plot the mean service time $\langle S_r \rangle$ as a function of resetting rate $r$ for different values of $\langle S_{on} \rangle$ fixing 
${\langle S \rangle}=1$ \text{~and~} $\alpha=1.5$. A minimum of $\langle S_r\rangle$ is obtained at an optimal resetting rate $r^*$ which varies with $\langle S_{on} \rangle$. Our next goal is to understand the behavior of mean queue length as a function of $CV_u$. To vary $CV_u$, we keep $\langle S \rangle=1$ and $\langle S_{on} \rangle=0.5$ fixed but modulate $\alpha$ in Eq.  (\ref{cv_deterministic_equation}). For each $\alpha$, we first optimize $\langle S_r \rangle$ as a function of $r$, compute the optimal resetting rate $r^*$, and then plugin the mean service time and the variability at this optimality into \eref{nr1}. Note that the existence of an optimal resetting rate $r^*$ is not guaranteed for any $CV_u$ as was discussed in Sec. (\ref{cvl1}). Clearly, for $r^*=0$, one finds $\langle N_{r^*}^I \rangle = \langle N \rangle$ and thus the plots for the underlying and reset process overlap with each other. However, as $r^*$ becomes finite, we can clearly see a deviation from the underlying PK formula and one observes $\langle N_{r^*}^I \rangle < \langle N \rangle$ as pointed out in Sec. \ref{cvl1}. The bottom curve in Fig. \ref{fig1}(b) shows how the mean queue length dramatically goes down as one gradually increases $CV_u$. 

\subsection{Case II: $CV_{on}=1$}
Next, we turn our attention to the case where $S_{on}$ is drawn from an exponential distribution, such that $f_{S_{on}}(t)=\gamma \exp(-\gamma t)$ where $\gamma>0$
is the rate parameter of the distribution. This implies $\langle S_{on} \rangle = \frac{1}{\gamma}$ and $\sigma^2(S_{on})=\frac{1}{\gamma^2}$ so that $CV_{on}=1$.  Therefore, the $CV^2_u$ of the
underlying process can be expressed as
\begin{align}
CV^2_u 
=\frac{e^{\alpha^2}-1+\frac{1}{\gamma^2}}{(1+ \frac{1}{\gamma})^2}.
\label{cv_exponential_equation}
\end{align}
As before, we keep $\langle S_{on} \rangle$ fixed at 0.5 and choose $\alpha$ to be the control parameter to vary $CV_u$. Similar optimization procedure is followed to obtain $r^*$ for different $\alpha$ and then is finally substituted into \eref{nr2}. For $CV_u<1$, we find $\langle N_{r^*}^{II} \rangle = \langle N \rangle$. However, as we increase $CV_u$, a clear deviation i.e., $\langle N_{r^*}^{II} \rangle < \langle N \rangle$ is observed. Here, $CV_u=1$ behaves as a sharp boundary where the transition occurs. The middle curve in Fig. \ref{fig1}(b) summarizes this behavior.

\subsection{Case III: $CV_{on}>1$}
Finally, to demonstrate the effect of resetting for $CV_{on}>1$, we consider the Weibull distribution $f_{S_{on}}(t)=\frac{k}{\nu}(\frac{t}{\nu})^{k-1} \exp\left(-\frac{t}{\nu}\right)^k$, $t \geq 0$ where $k > 0$ is the shape parameter and $\lambda > 0$ is the scale parameter of the distribution. In this case, the mean and variance of the overhead time are given by

\bea
\langle S_{on} \rangle=\nu \Gamma(1+\frac{1}{k}), \\
\sigma^2(S_{on})=\nu^2 \left[\Gamma \left(1+\frac{2}{k} \right) -\left( \Gamma \left(1+\frac{1}{k} \right) \right)^2 \right],
\eea
where $\Gamma(z)$ is the Gamma function of order $z$. 
 If $k$ is set to $0.7$, $\langle S_{on} \rangle$ becomes $1.2658\nu$ and thus
$CV_{on}=1.4624>1$ becomes independent of $\nu$. The variability of the
the underlying process can then be expressed as
\begin{align}
CV^2_u=\frac{e^{\alpha^2}-1+3.4268\nu^2}{(1+1.2658\nu)^2}.
\label{cv_weibull_equation}
\end{align}
We choose $\nu=0.395$ so that $\langle S_{on} \rangle$ is fixed at $0.5$. Performing the same optimization procedure, keeping $\langle S\rangle$ and $\langle S_{on} \rangle$ fixed, as done in the previous subsections, we can show $\langle N_{r^*}^{III} \rangle < \langle N \rangle$ when the condition (\ref{cvon_cr}) is satisfied. As also evident from Fig. \ref{fig1}(b), the deviation from the PK formula occurs only at a higher $CV_u>1$ in this case.

\begin{figure}[t]
\centering
\includegraphics[scale=0.7]{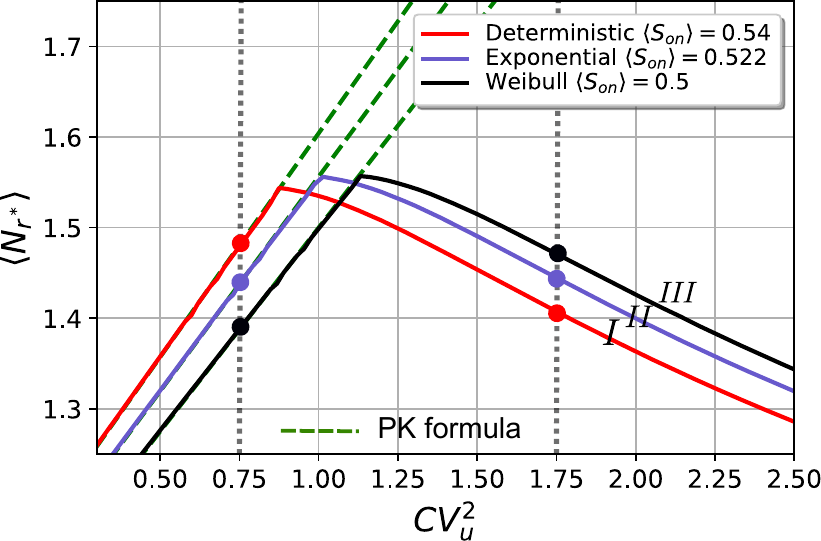}
\caption{Mean queue length at the optimal Poisson resetting as a function of squared $CV_u$ for different $\langle  S_{on} \rangle$ (as mentioned in the plot) and variability drawn from deterministic ($CV_{on}=0$), exponential ($CV_{on}=1$) and Weibull ($CV_{on}=1.4624$) distributions respectively. 
The underlying service time is drawn from log-normal distribution, whose density is given by \eref{log normal dist}. Here, we set $\langle S \rangle=1$ and vary $\alpha$ to control $CV_u$ via \eref{cv_deterministic_equation}, \eref{cv_exponential_equation}  and \eref{cv_weibull_equation} respectively. 
The dashed slanted lines indicate the 
Pollaczek-Khinchin formula in the absence of resetting. The left vertical dashed line with 
$CV_u<1$ indicates the order 
$    \langle N_{r^*=0}^I \rangle> \langle N_{r^*=0}^{II} \rangle>\langle N_{r^*=0}^{III} \rangle$ since $\langle S_{on}^I \rangle>\langle S_{on}^{II} \rangle>\langle S_{on}^{III} \rangle$ and thus resetting is seen to have no impact on the queue length. However, as soon as the optimal resetting rate becomes finite, we see a deviation from the PK formula and the order of the curves is reversed as can be seen from the right vertical dashed line with $CV_u>1$. In this case, we observe the order $\langle N_{r^*}^I \rangle< \langle N_{r^*}^{II} \rangle < \langle N_{r^*}^{III} \rangle$ for a given $CV_u$. Thus, optimally conducted resetting is seen to have more pronounced effect on the queues with smaller fluctuations in the overhead time albeit having a larger $\langle S_{on} \rangle$ compared to the queues with larger fluctuations and smaller mean.}
\label{fig2}
\end{figure}

Thus, resetting has more pronounced effect on the queue that experiences larger fluctuations in the overhead time albeit having a smaller $\langle S_{on} \rangle $.

\subsection{Mean queue length for different $\langle S_{on}\rangle$ }
So far we have assumed different $CV_{on}$ while keeping $\langle S_{on}\rangle$ fixed to estimate the mean queue length. Here, we relax this condition and study a combined effect when both of them are varied. Recall from the PK-formula that mean queue length increases linearly with $CV_u^2$ when optimal resetting rate is fixed at zero so that $\langle N_{r^*} \rangle = \langle N \rangle$. Similarly, for a fixed $CV_u$, one would expect that the length of the queue should be proportional to $\langle S_{on}\rangle$ i.e., for a large overhead time, the queue will also be longer. For instance, take a fixed $CV_u=\sqrt{0.75}$ in \fref{fig2} and vary $\langle S_{on} \rangle$. It is evident from the left vertical dashed line that the mean queue length increases  with an increasing $\langle S_{on} \rangle$. The coloured circles represent the respective values of the queue length for the cases with different $S_{on}$ as shown in \fref{fig2}. Thus, here, one has
\begin{align}
    \langle N_{r^*=0}^I \rangle> \langle N_{r^*=0}^{II} \rangle>\langle N_{r^*=0}^{III} \rangle.
\end{align}

\noindent
\textit{How can we compare between the queue lengths for different  $\langle S_{on} \rangle $ under optimal Poisson resetting?} It turns out that the mean length of a queue with a higher $ \langle S_{on} \rangle$
can be reduced more dramatically with the introduction of resetting compared to an another queue with a lower $ \langle S_{on} \rangle$ if the former has a lower $CV_{on}$. This can be seen again from \fref{fig2} as we set $CV_u=\sqrt{1.75}$ for which $r^*$ is non-zero and thus we are in the resetting-dominated regime. We can immediately note that the change in the order of the curves for $\langle N_{r^*} \rangle$ namely 
\begin{align}
    \langle N_{r^*>0}^I \rangle< \langle N_{r^*>0}^{II} \rangle < \langle N_{r^*>0}^{III} \rangle,
\end{align}
where the following order for the variability $CV_{on}^I<CV_{on}^{II}<CV_{on}^{III}$ and the mean $\langle S_{on}^I \rangle>\langle S_{on}^{II} \rangle>\langle S_{on}^{III} \rangle$ is maintained. In \fref{fig2}, we provide technical details that lead to this observation. While here we make this observation for the log-normal service time distribution, we believe that similar conclusions should hold for other service time distributions.

\section{Discussion and Summary} \label{Discussion}
Designing strategies that can optimize the number of jobs in a queue is an integral part of queuing theory. In particular, the key issue is to harness the large stochastic fluctuations in service times that can have deleterious effects in the performance of a queue. This has been alluded on various occasions in queuing theory in the context of computing workloads where the service time distributions have high variability. One such example arises in the UNIX process lifetime measurements \cite{Mor-book,Task-A-1}. In this perspective article, we review an interesting recent development which aims to address performance improvement of systems with high-variability workloads (see \cite{bonomo_pnas_2022,bonomo2023} and also \cite{queue-input}). We show that \textit{service resetting} can be a useful strategy to mitigate these problems. In particular, we consider a M/G/1 queue system where the jobs arrive at a constant rate and the server has two components: its own service time accompanied by an overhead/buffer time. The service is intermittently subjected to resetting and we have studied the ramifications of resetting protocols on the performance of the queue. We develop a renewal theory for the service time under resetting with overhead time and show how the modified service can be incorporated into the famous Pollaczek-Khinchin formula that provides an estimation of the mean length of the mean number of jobs in the queue.

Our analysis unveils three possible scenarios for the overhead time distribution: narrowly dispersed, marginally dispersed and broadly dispersed. In all these cases, we show that an optimally engineered resetting mechanism can either match or outperform the efficiency of the queue than the one without resetting. Specifically, we have shown that resetting
can dramatically reduce queue lengths when applied to servers that have high variability in the underlying 
service time. As such, resetting can alleviate the detrimental affect of large fluctuations by not only shortening the mean service time but also
reducing the relative stochastic fluctuations around this mean, hence
providing a two-fold advantage. 

\begin{figure}[t]
\centering
    \includegraphics[scale=1.02]{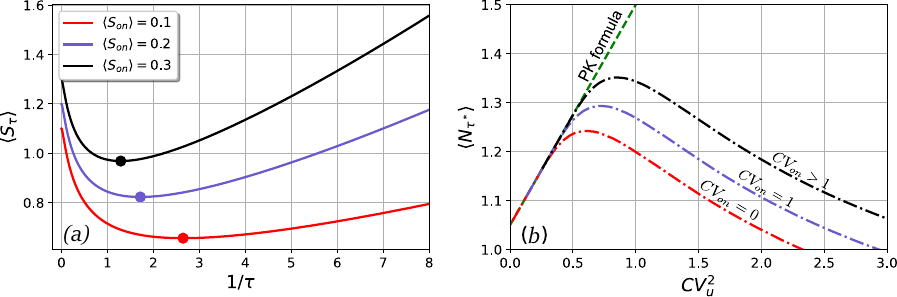}
     \caption{ Panel (a):  The mean service time from \eref{mean_sharp} as a function of $\tau$ for different overhead time $S_{on}$. The service time $S$ is taken from the log-normal distribution whose density is given by \eref{log normal dist}.  In each case, the optimal resetting rate $\tau^*$ can be identified by the solid circle, where $\langle S_{\tau} \rangle$ attains the global minimum. Here, we set $\langle S \rangle=1$, $\alpha=1.5$ and $\lambda=0.4$. Panel (b) shows $\langle N_{\tau^*} \rangle$ as a function of the underlying $CV_u^2$ (which varies as we change the control parameter $\alpha$). The Pollaczek-Khinchin formula gives the usual linear variation  (\eref{sharp-PK}) 
     as long as $\tau^*=0$. However, as soon as $\tau^*>0$, mean queue length gradually shortens and we see a deviation from that of the underlying process indicating the advantage gained by sharp resetting. Here, the overhead time $S_{on}$ is drawn from three distinct distributions with different $CV_{on}$ as mentioned in the plot.}
    \label{fig3}
\end{figure}

In the main text, we have only discussed the effects of Poissonian resetting. However, the general formalism developed therein can also be used to investigate other resetting strategy namely sharp or deterministic resetting \cite{PalJphysA,PalReuveniPRL17,Bhat-sharp}. Here, resetting is being conducted stroboscopically after a fixed time. This is strongly motivated from the earlier studies in the resetting community where sharp resetting has been proven to be a dominant strategy 
within the vast space of
stochastic restart strategies irrespective of the underlying process that is being restarted. To see the effect of this resetting strategy, we first assume that the resetting occurs after every $\tau$ units of time so that 
\begin{align}
    f_R(t)=\delta(t-\tau).
\end{align}
Using Eqs. (\ref{first-mom}) and (\ref{sec-mom}), one can immediately obtain the the mean and the second moment of the service time under sharp resetting \eref{appd} 
{\small\begin{align}
    \langle S_{\tau} \rangle&=  \frac{\int_{0}^{\tau}q_{S}(t)dt +\langle S_{on} \rangle}{1-q_S(\tau)} \label{mean_sharp}\\ 
     \langle S_{\tau}^2 \rangle &=
   \frac{2\int_0^{\tau}tq_S(t)dt +2 \langle S_{on}\rangle \int_{0}^{\tau}q_{S}(t)dt + \langle S_{on}^2 \rangle  }{1-q_S(\tau)} +\frac{2q_S(\tau)(\tau +\langle S_{on} \rangle)\left(\int_0^{\tau}q_S(t)dt+\langle S_{on}\rangle\right)}{(1-q_S(\tau))^2},
   \label{mom_sharp}
\end{align}}where $q_S(t)=\int_{t}^{\infty}dt'f_S(t')$ is the survival probability associated with the service. In simple words, it estimates the probability that the service has not occurred till time $t$ i.e., $q_S(t)=Pr(S>t)$. For the log-normal service time as defined in Sec. \ref{Application}, the survival probability $q_S(t)$ can be computed
\begin{align}
    q_S(t)=\frac{1}{2} \left[1+\text{erf}\left(\frac{\mu -\log (t)}{\sqrt{2} \alpha }\right)\right].
\end{align}
The mean queue length under sharp resetting can be obtained by substituting the metrics of the modified service time into  \eref{PK-1-r} 
\begin{align}
    \langle N_{\tau} \rangle=\frac{\rho_{\tau}}{1-\rho_{\tau}}+\frac{\rho_{\tau}^2}{2(1-\rho_{\tau})}\left( CV_{\tau}^2-1  \right),
    \label{sharp-PK}
\end{align}
where $\rho_\tau=\lambda \langle S_\tau \rangle$ and $CV_\tau$ is the variability of the service time under resetting. As before, we can find the optimal resetting time $\tau^*$ by setting $d \langle S_\tau \rangle/d\tau|_{\tau=\tau^*}=0$. Combining with this optimal relation, we plot the mean queue length $\langle N_{\tau^*} \rangle$ as a function of the underlying service time variability $CV_u$ for different overhead time distributions in \fref{fig3}(b). It is seen that resetting can reduce the number of jobs in a queue regardless of the specific choice for $CV_{on}$.

\begin{figure}[t]
\centering
    \includegraphics[scale=0.7]{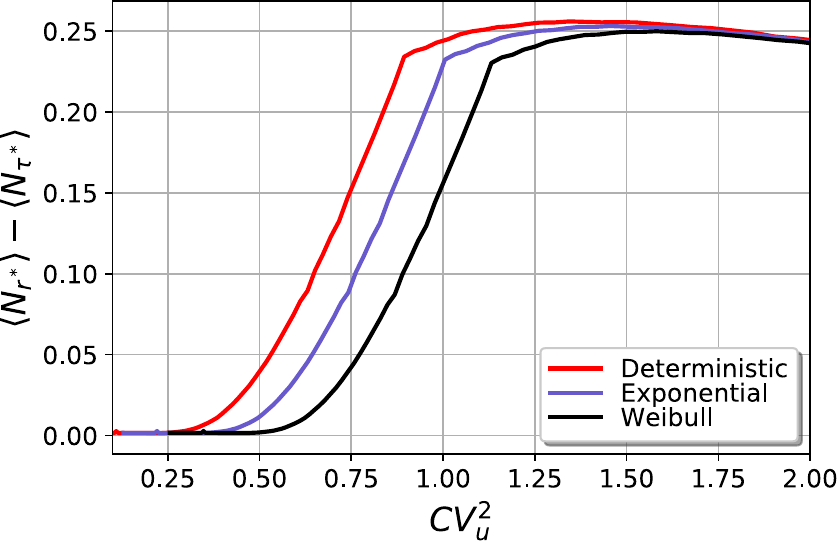}
    \caption{The difference between the mean queue length under Poisson optimal resetting and sharp optimal resetting as a function of squared $CV_u$. The service $S$ is taken from the log-normal distribution (density is given in \eref{log normal dist}) where we fix $\langle S \rangle=1$, and $\langle S_{on}\rangle=0.5$ while $\alpha$ is kept as the control parameter for $CV_u$. The overhead times $S_{on}$ are taken from different distributions: deterministic with $CV_{on}=0$, exponential with $CV_{on}=1$ and Weibull with $CV_{on}>1$. In each case, the difference $\langle N_{r^*} \rangle-\langle N_{\tau^*} \rangle$ is found to be positive conferring that optimal sharp resetting performs better than the optimal Poisson resetting. In this plot, we have set $\lambda=0.4$.}
    \label{fig4}
\end{figure}

It is moreover interesting to compare the mean queue length under optimal Poisson and sharp resetting.
To this end, we first recall from the theory of first passage under resetting that the mean completion time under optimal sharp resetting is always smaller or equal than that obtained under optimal Poissonian resetting. Here too, we find that the mean service time with overheads also respects the same relation (see \fref{fig3}(a)). Finally, we plot the difference between two optimally restarted queue lengths namely $\langle N_{r^*} \rangle-\langle N_{\tau^*} \rangle$ as a function of the underlying service time variability $CV_u$ for overhead times with $CV_{on} \{<1,=1,>1 \}$ in \fref{fig4}. Quite remarkably, we find that this difference strictly stays positive which essentially implies that the mean number of jobs in the
queue can be reduced further by resetting service at an optimal time rather than at an optimal rate. This observation is of practical importance since it reveals additional benefits in the performance modeling that can be gained by applying sharp resetting.

The formalism developed herein for the service time with overheads and resetting is not restricted to the M/G/1 queue, but in principle, can also be applied to other queues such as the G/G/1 queue where the arrivals are not necessarily Markovian \cite{Mor-book}. Similar renewal methods can also be employed to analyze queues when  the service time of the job has two components; one intrinsic component is the server slowdown and job's inherent size being the other extrinsic component \cite{bonomo2023}. Similar to \cite{bonomo_pnas_2022} and as shown here, resetting was shown to be a useful protocol to reduce the mean queue length. It will be interesting to study different trade-offs due to the overhead times in the above-mentioned queuing set-ups. There are many open frontiers with respect to the resetting based task assignment policies. Take for example a M/G/n queue that consists of a single queue but $n$ servers. When a server completes a job, it takes up the next job that is available at the head of the queue \cite{Mor-book}. It is possible to apply resetting protocols independently  to the individual servers as long as the `modified service' is applied to the same job. However, in a more naturalistic scenario, it is possible that a fraction of servers needs to be reset simultaneously. This renders a correlated queuing system and one can not readily apply the formalism developed here. These possible extensions to multiserver
queues, farms and networks are open for future research. 

Concluding, we believe that this perspective has 
shed light on the feasibility of applying resetting based strategies to the queuing systems. Hopefully, this will bridge a gap between the queuing and the resetting community and also will encourage the researchers to attempt and design more resetting based solutions in queuing systems with potential applications to computer science, randomized numerical algorithms,
and active living systems.

\section{Acknowledgements}
The numerical calculations
reported in this work were carried out on the IMSc-1 cluster, which is maintained and supported by the Institute of Mathematical Science’s High-Performance Computing Center. RR gratefully acknowledges support from the IMSc Post-Doctoral Fellowship. AP acknowledges research support from the Department of Science and Technology, India, SERB Start-up Research Grant Number
SRG/2022/000080 and Department of Atomic Energy,
Government of India.

\appendix
\section{Moment generating function of $S_R$}
\label{appa}
In this section, we try to find the general expression for the mean and second moment as written in \eref{first-mom} and \eref{sec-mom}. For that we first find the general Laplace transform or the moment generating function of $S_R$. Using that one can find all the moments in systematic fashion as we show below. We start by recalling \eref{renewal-1-main} from the main text
\begin{equation}
\begin{array}{l}
S_{R}=S_{on}+\left\{ \begin{array}{lll}
S &  & \text{if ~~}S<R\text{ ,}\\
 &\\
R+S_R'&  & \text{if~~ }R\leq S\text{ .}
\end{array}\right.\text{ }\end{array}
\end{equation} 
The above equation can also be written in a more compact way as
\begin{align}
    S_R=S_{on}+I(S<R)S+I(R\le S))R + I(R\le S) S_R',
    \label{ren2}
\end{align} 
where $I(R \le S)$ is the indicator function which takes the value 1 when $R\le S$ and zero otherwise. Thus,
\begin{align}
    \langle I(R \leq S) \rangle=Pr(R \leq S).
\end{align}
Let us now define $\widetilde{Z}(p)=\langle e^{-p Z} \rangle=\int f_Z(z) e^{-pz}dz$ as the moment generating function of the random variable $Z$, from which all its moments can be easily found. The Laplace transform of \eref{ren2}  can be given as
\begin{align}
   \widetilde{S}_R(p)&=\langle e^{-p S_R} \rangle \nonumber\\
    &=\langle e^{-p [S_{on}+I(S<R)S+I(R\le S))R + I(R\le S) S_R']} \rangle \nonumber\\
    &=\langle e^{-p S_{on}} \rangle \left[\langle I(R\le S) e^{-p S_{min}} \rangle +\langle I(S<R) e^{-p R_{min}-pS_R'} \rangle\right],
\end{align}
where $S_{min}\equiv \{S|S<R\}=$ and $R_{min}\equiv \{R|R<R\}$. We have also used the fact that $S_R'$ is an independent and identically distributed copy of $S_R$ and thus independent of $R~ \&~ S$. Performing the expectations over the indicator functions, we find
\begin{align}
    \widetilde{S}_R(p)&= \langle e^{-p S_{on}} \rangle \left[ Pr(S<R)\langle  e^{-p S_{min}} \rangle + Pr(R\le S)\langle  e^{-p R_{min}}  \rangle \langle  e^{-p S_R'}  \rangle \right]\nonumber\\
    &=\widetilde{S}_{on}(p) \left[ Pr(S<R)\widetilde{S}_{min}(p) + Pr(R\le S)\widetilde{R}_{min}(p) \langle  e^{-p S_R}  \rangle \right]\nonumber\\
    &=\widetilde{S}_{on}(p)\left[ Pr(S<R)\widetilde{S}_{min}(p) + Pr(R\le S)\widetilde{R}_{min}(p) \widetilde{S}_R(p) \right], 
    \end{align}
from where one finds
\begin{align}
  \widetilde{S}_R(p)&= \frac{Pr(S<R)\widetilde{S}_{min}(p) \widetilde{S}_{on}(p)  }{1-Pr(R\le S)\widetilde{R}_{min}(p)\widetilde{S}_{on}(p) }.
\end{align}
This is an exact expression for the distribution of $S_R$ in Laplace space. This is also the moment generating function from which $n^{th}$ moment of $S_R$ can be computed directly via
\begin{align}
    \langle S_R^n \rangle=(-1)^n\left.\frac{d^n \widetilde{S}_R(p)}{dp^n}\right|_{p=0}.
    \label{srn}
\end{align}
For instance, the first moment reads
{\small
\begin{align}
    \langle S_R \rangle&=-\left.\frac{d\widetilde{S}_R(p)}{dp}\right|_{p=0}\nonumber\\
    &= \frac{Pr(S<R)\left[(1-Pr(R\le S))(\langle S_{on}\rangle +\langle S_{min} \rangle)+Pr(R\le S) \langle S_{on}\rangle+ Pr(R\le S)\langle R_{min} \rangle\right]}{(1-Pr(R\le S))^2} \nonumber\\
    &= \frac{Pr(S<R)\left[ Pr(S<R)\langle S_{min} \rangle + Pr(R\le S)\langle R_{min} \rangle +\langle S_{on}\rangle\right]}{(Pr(S<R))^2} \nonumber\\
    &=\frac{ \langle \text{min}(S,R) \rangle + \langle S_{on}\rangle}{Pr(S<R)},
\end{align}}which is \eref{first-mom} with the identification $\langle \text{min}(S,R) \rangle=Pr(S<R) \langle S_{min} \rangle +Pr(R\le S)\langle R_{min} \rangle$. A similar exercise for the second moment gives
{\begin{align}
    \langle S_R^2 \rangle&=\left.\frac{d^2\widetilde{S}_R(p)}{dp^2}\right|_{p=0}\nonumber\\
    &=\frac{\langle (\text{min}(S,R)+S_{on})^2\rangle }{Pr(S<R)}+\frac{2Pr(R\le S)(\langle R_{min}\rangle +\langle S_{on} \rangle)\left(\langle \text{min}(S,R) \rangle + \langle S_{on}\rangle\right)}{Pr(S<R)^2}.
\end{align}
which is identified  as the \eref{sec-mom} in the main text. \eref{srn} thus encodes all the information about the higher moments.

\section{Moments of the service time for Poissonian resetting}
\label{appb}
In this section we take the representative case when the resetting times are drawn from an exponential distribution given by
\begin{align}
    f_R(t)=re^{-rt}.
\end{align}
where $r$ is the resetting rate.
The cumulative function is given by
\begin{align}
    Pr(R \leq t)=1-e^{-rt}.
\end{align}
The distribution of the random variable $\text{min}(S, R)$ can be computed by noting
\begin{align}
Pr(\text{min}(S, R) \leq t) &= 1- Pr(\text{min}(S, R) > t) \nonumber\\
&=1 - Pr(S > t)Pr(R > t) \nonumber\\
&= 1 - Pr(S > t) e^{-rt},
\label{cumulative}
    \end{align}
from which one can gets
\bea
f_{min(S,R)}(t) = e^{-rt} f_S(t)  + r e^{-rt} Pr(S > t) ,
\label{min-PDF}
\eea
which can be used to compute all the moments for the random variable  $\text{min}(S, R)$. 

\noindent
\textbf{Mean service time with overheads:} To compute the mean, we note that the first term on the numerator in \eref{first-mom}) can be written as
\begin{align}
\langle \text{min}(S,R) \rangle 
     &=\frac{1-\widetilde{S}(r)}{r},
\end{align}
where $\widetilde{S}(r)=\int_0^\infty dt~e^{-st}~f_S(t)$. The denominator can be written as
\begin{align}
    & Pr(S<R)=\int_0^{\infty}dt ~ f_S(t) Pr(R>t).
\end{align}
For exponential resetting times $Pr(R>t)=e^{-rt}$ and we have
\begin{align}
    Pr(S<R)&=\int_0^{\infty}dt~ f_S(t) e^{- r t} =\widetilde{S}(r).
\end{align}
Combining together, we find
\begin{align}
 \langle S_r \rangle &= \frac{1-\widetilde{S}(r)+r \langle S_{on} \rangle}{r \widetilde{S}(r)}    ,
\end{align}
which is \eref{mean under restart} in main text.

\noindent
\textbf{Second moment of the service time with overheads:} To find the second moment for Poisson resetting, we now use \eref{sec-mom}. The expectation  $\langle \text{min}(S,R)^2 \rangle$ in the numerator can be computed directly using \eref{min-PDF} 
\begin{align}
\langle \text{min}(S,R)^2 \rangle &=\int_0^{\infty} dt~ t^2 \left( f_S(t)Pr(R>t)+f_R(t)Pr(S>t) \right) \nonumber\\
&=\int_0^{\infty} dt~ t^2 \left( f_S(t)~e^{-rt} +r~Pr(S>t)~e^{-rt} \right) \nn \\ 
&= \int_0^{\infty} dt~ t^2 f_S(t)~e^{-rt} + r\int_0^{\infty} dt~ t^2~Pr(S>t)~e^{-rt} \nn \\ 
&= \frac{d^2\tilde{S}(r)}{dr^2} + r\int_0^{\infty} dt~ t^2 \left( 1-Pr(S<t)\right)e^{-rt} \nn \\
&= \frac{d^2\tilde{S}(r)}{dr^2} +\frac{2}{r^2} - r\int_0^{\infty} dt~ t^2Pr(S<t)e^{-rt} \nn \\
&= \frac{d^2\tilde{S}(r)}{dr^2} +\frac{2}{r^2} - r \frac{d^2 \left (\frac{ \tilde{S}(r)}{r}\right )}{dr^2} \nn \\
&= \frac{2r~\frac{d\tilde{S}(r)}{dr} -2\tilde{S}(r)+2}{r^2}. \label{second mom min SR Poiss}
\end{align}
In the above derivation we have used the following property $\int_0^\infty dt ~t^nf(t)e^{-rt}=(-1)^n\frac{d^n\tilde f(r)}{dr^n}$. Next, to calculate $\langle R_\text{min} \rangle $
we need to use the density of the conditional time $R_{min}$ which is given by
\begin{align}
    f_{R_{\text{min}}}(t) = \frac{f_R(t) \int_t^\infty~dt'~f_S(t')}{Pr(R\leq S)}=\frac{f_R(t) Pr(S>t)}{Pr(R\leq S)} \label{rmin}.
\end{align} 
Then, $\langle R_\text{min} \rangle $ can be expressed as
\begin{align}
    \langle R_\text{min} \rangle&=\int_0^{\infty} dt~ tf_{R_{min}}(t) \nn \\ 
&= \frac{1}{Pr(R\leq S)}\int_0^{\infty} dt~t~re^{-rt} Pr(S>t) \nn \\ 
&= \frac{r}{1-Pr(S<R)}\int_0^{\infty} dt~t~e^{-rt} \left(1-Pr(S<t)\right) \nn \\ 
&= \frac{r\left(\int_0^{\infty} dt~t~e^{-rt}-\int_0^{\infty} dt~t~e^{-rt} Pr(S<t)\right)}{1-\tilde{S}(r)} \nn \\ 
&=\frac{r\left(\frac{1}{r^2} + \frac{d\left(\frac{\tilde{S}(r)}{r}\right)}{dr}\right)}{1-\tilde{S}(r)} \nn \\
&= \frac{r~\frac{d\tilde{S}(r)}{dr} -\tilde{S}(r)+1}{r\left(1-\tilde{S}(r) \right)}. \label{Rmin Poiss}
\end{align}
 Substituting Eqs. (\ref{second mom min SR Poiss}) and (\ref{Rmin Poiss}) into \eref{sec-mom}, one obtains 
\begin{align}
    \langle S_r^2 \rangle=\frac{2 r \frac{d \widetilde{S}(r)}{dr}(1+r \langle S_{on}\rangle)+2(1-\widetilde{S}(r))(1+r \langle S_{on} \rangle)^2+
r^2 \widetilde{S}(r)\langle {S_{on}^2} \rangle}{r^2 \widetilde{S}(r)^2},
\end{align}
which was announced in \eref{secmom under restart} in the main text.

\section{General discussion on the ``resetting induced efficiency criterion'' for the mean queue length with different variability $CV_{on}$}
\label{appc}
In this section, we elaborate more on the general 
conditions that were obtained in Sec. (\ref{Service at ORR}) to underpin the effect of resetting. We start by deriving the most general criterion that ensures the existence of an optimal $r^*$. To see this, we introduce an infinitesimal resetting rate $\delta r$ and ask under what condition the following inequality $ \langle S_{\delta r \to 0} \rangle<\langle S_u \rangle$
holds. Expanding \eref{mean under restart} in the power of $\delta r$ and imposing the above condition, we derive a universal relation
\begin{align}
     CV^2>1+2\frac{\langle S_{on} \rangle}{\langle S \rangle},
     \end{align}
that guarantees the existence for an optimal resetting rate \cite{IP}. In terms of $CV_u$ with the help of \eref{cvu} this criterion takes the form
\begin{align}
    CV^2_u>1+ \frac{\langle S_{on} \rangle^2}{\langle S_u \rangle^2}(CV_{on}^2-1).
    \label{cvu_crt}
\end{align}
The above equation will be central to our remaining discussion where we study various cases for $CV_{on}$.

\subsection{Case I: $CV_{on}<1$}
\label{cvl1}
In \eref{con1} in the main text we argued that for $CV_u>1$ the mean queue length can be reduced by introducing resetting when $CV_{on}<1$. However, we emphasize that this is not a necessary (albeit sufficient) condition. From \eref{cvu_crt} we notice that the RHS is less than one for $CV_{on}<1$. This in turn implies a finite optimal value of $r^*$ also can be found under the same condition. This implies that $\rho_{r^*}$ becomes less than $\rho$ for  $CV_u<1$. As a result, for a sufficiently small  $\rho_{r^*}$, one can still have $ \langle N_{r^*}^{I} \rangle < \langle N \rangle$ from \eref{nr1}. In \fref{fig2} we indeed find that the deviation occurs at value of $CV_u$ which is less than unity.

\subsection{Case II: $CV_{on}=1$}
\label{cv1}
In this case the RHS of \eref{cvu_crt} becomes exactly unity and hence, a finite optimal $r^*$ shows up
only when $CV_u>1$. It is thus evident from
\eref{con2} that $CV_u>1$ is both a necessary and sufficient condition for the reduction in the mean queue length. As shown in \fref{fig2}, the transition point (where $ \langle N_{r^*}^{II} \rangle < \langle N \rangle$) is exactly found at $CV_u=1$.

\subsection{Case III: $CV_{on}>1$}
\label{cvg1}
In this case the RHS of \eref{cvu_crt} is strictly greater than one. One can thus expect to find an optimal $r^*$ for values of $CV_u>1$. Thus the condition $ \langle N_{r^*}^{II} \rangle < \langle N \rangle$ will be satisfied only when $CV_u>1$. This is also evident from \fref{fig2} where we see that the transition point occurs at $CV_u>1$.

To derive the criterion (\eref{cvon_cr}) as mentioned in the main text we recall the PK formula from \eref{nr3} for the optimally restarted process and the same without resetting from \eref{PK-1}
\begin{align}
    \langle N_{r^*}^{III} \rangle&= \frac{\rho_{r^*}}{1-\rho_{r^*}}+\frac{\lambda^2}{2(1-\rho_{r^*})} \frac{\langle S_{on}\rangle^2}{\tilde S(r^*)}\left( CV_{on}^2-1  \right) \\
   \langle N \rangle  & = \frac{\rho}{1-\rho}+\frac{\rho^2}{2(1-\rho)}\left( CV_u^2-1  \right),   
\end{align}
and impose the condition $ \langle N_{r^*}^{III} \rangle<\langle N \rangle$. For a finite $r^*$, this would yield
\begin{align}
   &\lambda ^2\frac{\langle S_{on}\rangle^2}{\tilde S(r^*)}\left( CV_{on}^2-1  \right)< \rho^2\left( CV_u^2-1  \right)\nonumber\\
   &CV_u^2>1+\left(\frac{\langle S_{on}\rangle}{\langle S_u \rangle}\right)^2\frac{1}{\Tilde{S}(r^*)}(CV_{on}^2-1),
   \label{assm}
\end{align}
which gives the criterion obtained in \eref{cvon_cr} where we have substituted $\rho=\lambda\langle S_u \rangle$. Finally, we remark that this is a sufficient condition (but not necessary). In other words, one can still find a $CV_u$ where this condition is not satisfied but the inequality $ \langle N_{r^*}^{III} \rangle<\langle N \rangle$ holds.

\section{Moments for sharp resetting times}
\label{appd}
This section is dedicated to compute the first two moments of the service time under sharp resetting. Here, resetting occurs always after a fixed time interval $\tau$ such that
\begin{align}
    f_R(t)=\delta(t-\tau),
\end{align}
from which one can see
\begin{align}
    Pr(R>t)=\theta(\tau-t),
\end{align}
where $\theta(z)$ is the Heaviside theta function which takes value unity only when $z>0$ and zero otherwise.

\noindent
\textbf{Mean service time with overheads: } The numerator in \eref{first-mom} can be computed in the following way
\begin{align}
     \langle \text{min}(S,R) \rangle&=\int_{0}^{\infty}Pr(S>t)Pr(R>t) dt \nonumber\\
     &=\int_{0}^{\infty}Pr(S>t)\theta(\tau-t) dt\nonumber\\
     &=\int_{0}^{\tau}dt ~q_{S}(t), \label{Sharp numerator}
\end{align}
where $q_{S}(t)=Pr(S>t)$ is the survival probability of the process $S$ up to time $t$, which indicates the probability that the service has not yet been completed up to time $t$. The denominator in \eref{first-mom} is found to be
\begin{align}
    Pr(S<R) &=\int_0^\infty~dt~f_R(t)Pr(S<t) \nn \\
 &=\int_0^\infty~dt~\delta(t-\tau)Pr(S<t) \nn \\
&= \int_0^\infty~dt~\delta(t-\tau)(1-q_S(t)) \nn \\
&= 1-q_S(\tau). \label{Sharp denominator}
\end{align}
Substituting \eref{Sharp numerator} and \eref{Sharp denominator} in \eref{first-mom} we obtain the first moment for sharp resetting as
\begin{align}
    \langle S_{\tau}  \rangle &= \frac{\int_{0}^{\tau}q_{S}(t)dt +\langle S_{on} \rangle}{1-q_S(\tau)},
\end{align}
which was announced in \eref{mean_sharp}.

\noindent
\textbf{Second moment of service time with overheads: } Following the same procedure as the previous section, we find 
\begin{align}
    \langle \text{min}(S,R)^2 \rangle&=\int_0^{\infty} dt~ t^2 \left( f_S(t)Pr(R>t)+f_R(t)Pr(S>t) \right) \nonumber\\
    &=\int_0^{\infty} dt~ t^2 \left[-\frac{\partial q_S(t)}{\partial t}\theta(\tau-t) + q_S(t)\delta(\tau-t) \right] \nn \\ &= \tau^2q_S(\tau) - \int_0^{\tau} dt~ t^2\frac{\partial q_S(t)}{\partial t} \nn \\ &= \tau^2q_S(\tau) -\tau^2q_S(\tau) + 2\int_0^{\tau} dt~ t~q_S(t) \nn \\ &= 2\int_0^{\tau} dt~ t~q_S(t).
\label{squared min}
\end{align}
The quantity $\langle R_{min} \rangle$ can also be obtained from \eref{rmin} directly
\begin{align}
\langle R_\text{min} \rangle&=\int_0^{\infty} dt~ tf_{R_{min}}(t) \nn \\ &= \frac{1}{Pr(\tau\leq S)}\int_0^{\infty} dt~ t\delta(t-\tau) Pr(S>t) \nn \\ &= \frac{\tau q_S(\tau)}{q_S(\tau)} \nonumber\\
&= \tau.
\label{Sharp Rmin}
\end{align}
Using Eqs. (\ref{squared min}) and (\ref{Sharp Rmin}) in \eref{sec-mom}, we arrive at the following expression for the second moment of service time with overheads
{\small\begin{align}
  \langle S_{\tau}^2 \rangle &=
   \frac{2\int_0^{\tau}tq_S(t)dt +2 \langle S_{on}\rangle \int_{0}^{\tau}q_{S}(t)dt + \langle S_{on}^2 \rangle  }{1-q_S(\tau)} +\frac{2q_S(\tau)(\tau +\langle S_{on} \rangle)\left(\int_0^{\tau}q_S(t)dt+\langle S_{on}\rangle\right)}{(1-q_S(\tau))^2},
\end{align}
which is \eref{mom_sharp} in the main text.

\end{document}